   \documentstyle[epsf]{mn}
  \newcommand{\Msolar}{\mbox{\,$\rm M_{\odot}$}}        % solar mass
  \newcommand{\Teff}{\mbox{\,\em T$_{\rm eff}$}}         % effective temperature
  \newcommand{\kmsec}{\,\mbox{$\mbox{km}\,\mbox{s}^{-1}$}}    % kilometres/second
  \def\simge{\mathrel{\raise1.16pt\hbox{$>$}\kern-7.0pt
    \lower3.06pt\hbox{{$\scriptstyle \sim$}}}}           % approx ge
  \def\simle{\mathrel{\raise1.16pt\hbox{$<$}\kern-7.0pt
    \lower3.06pt\hbox{{$\scriptstyle \sim$}}}}           % approx le
\begin{document}

    \title[Radial velocities of pulsating subdwarf B stars]
         {Radial velocities of pulsating 
          subdwarf B stars: KPD\,2109+4401 and PB\,8783
          \thanks{Based on observations
           obtained with the William Herschel Telescope, ING, La Palma,
           Spain} }

     \author[C.S.Jeffery \& D.Pollacco ]
         {C. Simon Jeffery$^{1}$
                    and
          Don Pollacco$^{2,3}$
          \\
          $^{1}$Armagh Observatory, College Hill, Armagh BT61 9DG, 
                Northern Ireland\\
          $^{2}$Isaac Newton Group, La Palma\\
          $^{3}$Queens University Belfast, Belfast BT7 1NN, Northern Ireland }

     \date{Accepted .
           Received ; }

     \pagerange{\pageref{firstpage}--\pageref{lastpage}}
%     \pubyear{1999}

\label{firstpage}

    \maketitle

%
% _____________________________________________________________________ abstract
%
\begin{abstract}

High-speed spectroscopy of two pulsating subdwarf B stars,
KPD\,2109+4401 and PB\,8783, is presented. Radial motions are detected with
the same frequencies as reported from photometric observations and with 
amplitudes of $\sim 2\kmsec$ in two or more independent modes.
These represent the first direct observations of surface motion due to 
multimode non-radial oscillations in subdwarf B stars.
In the case of
the sdB+F binary PB\,8783, the velocities of both components are resolved;
high-frequency oscillations are found only in the sdB star and not the F star.
There also appears to be evidence for mutual motion of the binary components.
If confirmed, it implies that the F-type companion is $\simge1.2$ times more 
massive than  the sdB star, while the amplitude of the F star acceleration 
over 4 hours would constrain the orbital period  to lie between 0.5 and 3.2d. 

\end{abstract}


% ________________________________________________________________ introduction
%

\section{Introduction}

The discovery of small-amplitude non-radial pulsations in a number of
subdwarf B stars (sdBVs) has introduced a powerful new tool for the study of
stellar remnants  (Kilkenny et al. 1997=SDBV\,I and subsequent papers). 
%% 1998=SDBV\,VIII, 1999=SDBV\,X,  
%% Koen et al. 1997=SDBV\,II, 1998a=SDBV\,VII, 1998b=SDBV\,XI, 
%% Koen 1998=SDBV\,IX,
%% Stobie et al. 1997=SDBV\,III, 
%% O'Donoghue et al. 1997=SDBV\,IV, 1998a=SDBV\,VI, 1999b=SDBV\,V, 
%% Bill\`eres et al. 1997, 1998). 
Long time-series photometric campaigns have detected rich spectra of
oscillations in over a dozen targets, with frequencies
and amplitudes generally indicative of low-order ($\ell=0-2$) 
and low-degree modes.  
The discovery has revolutionized the study of subdwarf B stars for the
simple reason that pulsations in these stars were not expected. 
From a theoretical point of view, it appears that the pulsation 
mechanism is  only effective when diffusion processes modify the outer 
layers of the star.  Metal-enrichment in a specific layer must conspire 
with the local temperature to drive pulsations through the opacity mechanism
(Charpinet et al. 1997). 

In terms of their effective temperature and surface gravity, sdBVs are
indistinguishable from their non-variable counterparts -- there is no
instability strip in which all sdBs pulsate. 
The origin of all sdBs 
remains a puzzle, although the increasing detection of sdB binarity,
including  several sdBVs, may point to a previous phase of
common-envelope evolution in many, if not all. 
The structure of sdBs is partially hidden as a consequence of
atmospheric diffusion, which disguises the true composition 
 and the mass of the hydrogen-rich envelope, both of
 which are key diagnostics of 
previous evolution. By enabling an exploration of the composition and
mass of these outer layers, asteroseismology may be of
a pivotal importance.

Immediately after the discovery of sdBV pulsations, we
recognized the potential for spectroscopy to provide additional diagnostics.
Mode 
identification from photometry alone is challenging, 
and could be assisted by the identification of
line-profile variations, whilst the comparison of radial and light
amplitudes could be used to determine stellar radii directly.
Spectroscopy might also demonstrate the presence of higher-order
modes not detected photometrically.
The limitations are that the periods are short (100-500s) compared 
with conventional CCD readout times, the stars are faint 
(12--15 mag.), and the photometric amplitudes are typically only
a few tenths of one per cent. 
High-resolution high-S/N multi-line studies such 
as those obtained for non-radial oscillations in rapidly rotating
bright O and B stars (e.g. Reid et al. 1993, Telting, Aerts \& Mathias 1997)
would not appear to be feasible.
However, the development of new techniques offered the possibility to 
acquire high-speed spectroscopy of sdBVs and here we report our 
first successful observations obtained in 1998 October. 
Subsequently, O'Toole et al. (2000) announced preliminary results of
independent radial velocity observations. They report a 9\kmsec\ amplitude
periodic variation in the large amplitude sdBV PG1605+072
at the principal frequency of 2.10 mHz found photometrically by Koen et al. 
(1998, SDBV\,VII). 

\section{Observations}

The targets selected for our initial study were PB\,8783 
(Koen et al. 1997 = SDBV\,II) and
KPD\,2109+4401 (Koen et al. 1998 = SDBV\,XI, Bill\`eres et al. 1998). 
Although not ideal, 
our observations were scheduled when possibly more
exciting targets such as PG1336--018 (Kilkenny et al. 1998 = SDBV\,VIII) 
and PG1605+072 (SDBV\,VII) were not accessible.
On the other hand, both targets 
have been well-studied photometrically (O'Donoghue et al 1998 = SDBV\,V, 
SDBV\,XI, Bill\`eres et al. 1998)
and show multiple periods in the ranges 7--10 and 5--6 mHz respectively, 
with amplitudes of a few millimagnitudes.

Observations were obtained with the 4.2m William Herschel Telescope
on 1998 October 3 and 4 using the blue arm of the intermediate
dispersion spectrograph ISIS. The R1200B grating and the TEK1 CCD
with $1124^2$ 24$\mu$m pixels
yielded an instrumental resolution (2 pixel) $R=5\,000$ in the
wavelength interval 4020--4420 \AA. This wavelength region was chosen
because it contains two strong Balmer lines and, potentially, 
a number of neutral helium and minor species lines that are normally
observed in early-B stars. It also maximises the photon collection rate.

%
%    Table                                                              Observation Summary
%
\begin{table}
\caption[WHT Observations]
{Summary of 1998 October WHT driftmode observations of sdBVs}
\label{sdb_obs}
\small
\begin{center}
\begin{tabular}{lc ccc ccc}
Star           & Date        &  V & $t$(s) & $c$ & $s$ & $j$ & $\Sigma n$   \\ [1mm]
PB\,8783       & 1998 Oct 3 &  12.3 &  7.9 & 325 & 23 & 57 & 1400         \\
KPD\,2109      & 1998 Oct 4 &  13.4 &  9.8 & 230 & 14 & 57 & 1200         \\
\,+4401        & & & & & & & \\
\end{tabular}
\end{center}
\end{table}

The CCD was read out in low-smear drift mode (Rutten et al. 1997), 
in which only 
a small number $j$ of CCD rows (parallel to the dispersion direction) 
are read out at one time. A dekker is used to limit the slit-length,
thus only a fraction of the CCD window, 
$\simeq j$ rows, is exposed at one time. After exposing for an interval
$t$, the CCD contents are stepped down by $j$ rows. This charge
transfer imposes a small dead time
$d$ which is roughly proportional to the CCD window size.
Each set of $j$ rows is accumulated into a data cube
containing $n$ individual 2D spectra, stacked adjacent to one another. 
In practice the first few spectra in the frame are null, corresponding 
to the unexposed part of the CCD being read out before exposed rows 
reach the CCD edge. Comparison lamps and stellar spectra can be obtained 
in exactly the same way. Maximum frame sizes limit $n$ to a few hundred;
we normally adopted $n=200$ before beginning a new frame. 

In our experiment, $j$ was set just wide enough 
to admit a small region of sky either side of the stellar image and
to allow for edge effects between adjacent spectra. On-chip binning by
a factor of 3 further reduced the read-out time and data-volume. 
$t$ was chosen so that 
$t+d$ was approximately 1/10 of the typical observed pulsation period. 
Values for $t, j$ and $\Sigma n$ are given for each target in Table~\ref{sdb_obs},
together with the average number of counts detected per wavelength resolution element
$c$, and the maximum S/N ratio $s$ anticipated in each exposure from nominal 
performance figures (SIGNAL, Benn 1997). In this configuration, $d=1.57$s 
and the read-out noise was 5.24 electrons per pixel.

Due to occasional errors in the CCD controller, not all exposure times $t$ 
were equal, roughly one in 100 exposures would drop by an unpredictable 
amount. These were identified by integrating the total counts in each 
individual spectrum and adjusting individual exposure times to reflect the
photometry. 
 
The data cubes were reduced using scripted standard IRAF
routines. After bias subtraction and flatfielding, the 1-dimensional
spectra were optimally extracted and sky subtracted. The copper-argon
arc spectra obtained at the beginning and end of each data cube were
time weighted for each stellar observation and the corrected
calibration derived and applied. Over the duty cycle of one data cube
the arc shifts were never greater than a few microns.  Further
reductions included normalization of the spectra and cosmic ray
removal.

Cross-correlation of the arc spectra showed that the maximum shift
during a half hour period was around 7 microns (about a quarter of a
pixel). Arc exposures showed that this was varying smoothly during the
night and as calibration data were obtained before and after each
stellar data cube, then simple spectral interpolation of the extracted
one dimensional spectra (weighted by the time interval from the
bracketing arc spectra) removed any instrumental shift to probably
within 1 micron ($<0.9\kmsec$).

%
% Figure                                                     Mean Spectra  
%
\begin{figure}
\epsfxsize=85mm
\epsfbox{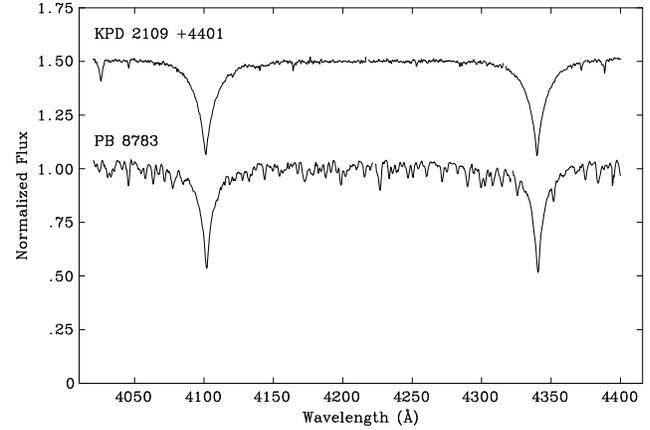}
\caption[Mean Spectra]{The mean spectrum obtained for each of two pulsating 
sdB stars observed in 1998 October. See text for further details.}   
\label{fig_means}
\end{figure}

The final data product is a set of files each containing $n$ 
1-dimensional wavelength-calibrated flux-normalized spectra. Each spectrum
is time-tagged. The mean spectrum for each of our targets is shown in 
Fig.~\ref{fig_means}

%
% Figure                                            KPD2109+4401 observations
%
\begin{figure}
\epsfxsize=85mm
\epsfbox{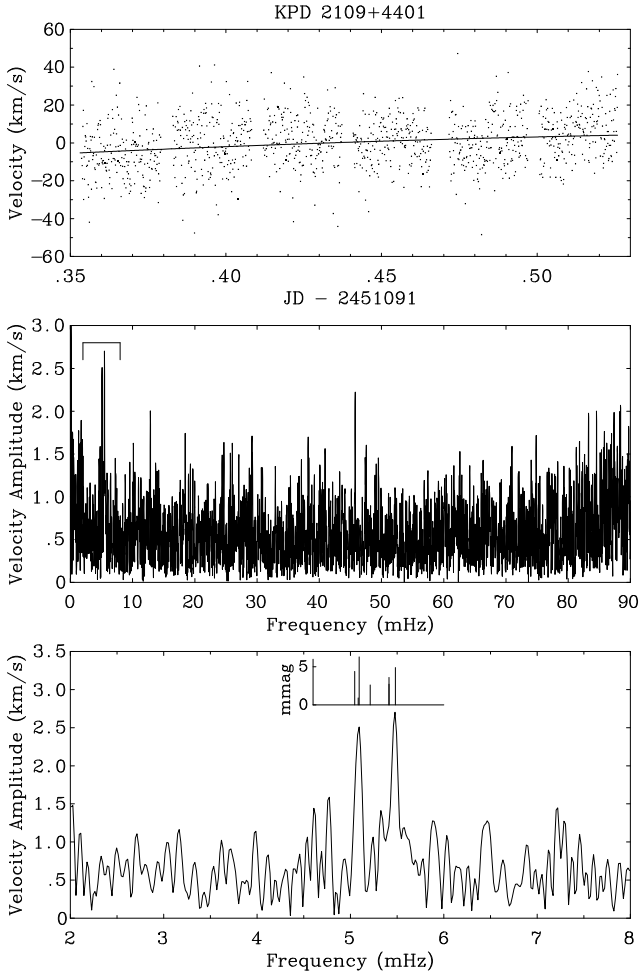}
\caption[KPD2109+4401]{Radial velocities and amplitude spectrum of KPD\,2109+4401
from WHT high-speed spectroscopy on 1998 Oct 4. 
The slowly varying function $b(t)$ is shown in the top panel. 
The bottom panel is an enlargement of the marked section of the 
amplitude spectrum shown in the middle panel. The
amplitudes of photometric frequencies inset in the bottom panel are
are taken from SDBV\,IX.}   
\label{fig_KPD2109}
\end{figure}

%
% Figure                                            PB8783 observations
%
\begin{figure}
\epsfxsize=85mm
\epsfbox{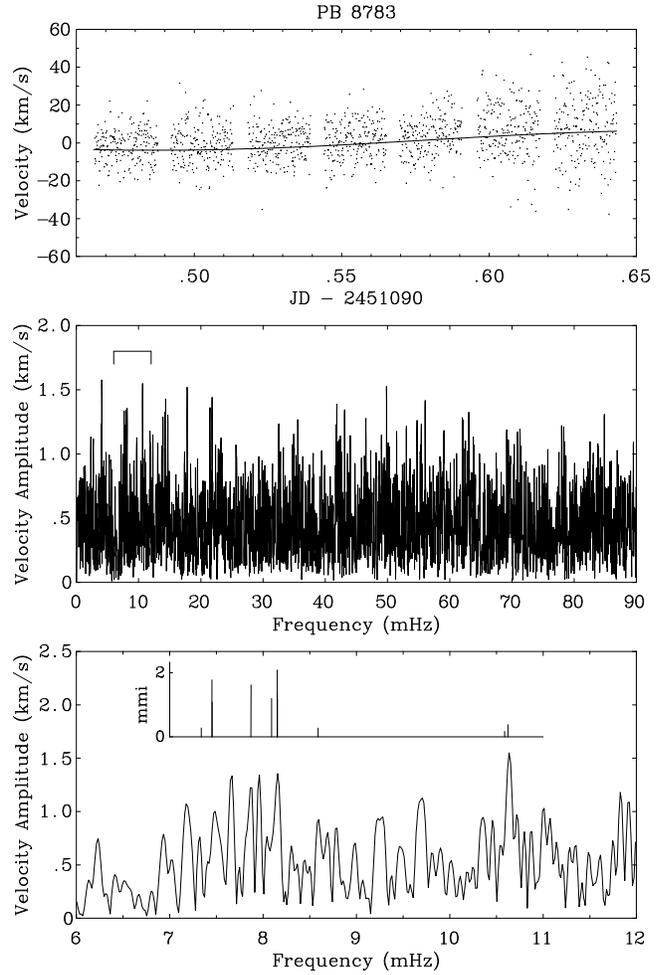}
\caption[PB8783]{As Fig.~\ref{fig_KPD2109} for PB\,8783 (1998 Oct 3). 
The photometric frequencies are taken from SDBV\,V,  
mmi denotes parts-per-thousand and 1 mmi equals 1.086 mmag. }   
\label{fig_PB8783}
\end{figure}

\section{Analysis and Results}

The primary objective of our observations was to detect and measure  
radial motion due to the dominant oscillatory modes detected from
photometry. There was the additional possibility that line-profile
variations or multiple modes might be detected, as well as any mutual motion
within any binary system. 

The most commonly adopted method for measuring radial-velocity shifts 
in stellar spectra is the cross-correlation function (ccf). We constructed a
mean spectrum for each target.  This template was cleaned to remove 
stationary features such as  bad CCD columns.  Both template and
individual spectra were continuum-subtracted; they were already linearized 
to the same wavelength scales. The ccf for
each spectrum relative to the template was computed and stored as a
2d function $\chi(t,v)$, where $t$ is now time, $\delta t$ is the mean sampling 
interval and $v$ is radial velocity
relative to the template.
The velocity of each spectrum was found by 
a) fitting a gaussian to, 
b) fitting a parabola to 
and c) computing the centroid of $\chi(v)$ in the range $-\delta v$ to $+\delta v$, 
to give functions $v_g(t), v_p(t),$ and $v_c(t)$, respectively. 
Moments
$M_j(t) = \int (v-v_{\rm ref})^j \chi(v,t) dv; j = 1,2,3$ were also computed.
With $v_{\rm ref}=0$, $M_1$ corresponds to the centroid.
$v(t)$ was sensitive both to the method chosen and the value of $\delta v$. 

The functions $v(t)$ were investigated for periodic content by means
of the discrete Fourier transform ${\cal F}$. 
Either a low-degree polynomial or a long-period sine function 
$b(t)$  was first fit and subtracted from $v(t)$. 
${\cal F} \{v(t)-b(t)\}$  was then 
computed.

The velocity data $v_p(t)$  and  $\cal F$ are
shown in Figs.~\ref{fig_KPD2109} and 
\ref{fig_PB8783}. These figures also show a 
representation of the principal frequencies and amplitudes 
discovered photometrically $\cal P$ (SDBV\,V,IX) \footnote{Throughout this
paper, the term amplitude applied to periodic signals refers to
the semi-amplitude $a$ of the sine function $a\sin(2\pi\nu t +\phi)$}.

The information we require from the amplitude spectrum 
is not in the first instance the significance of the peaks but
rather the amplitude associated with expected 
frequencies. 
Over the entire range 
$0.1\sim2/\Sigma n\delta t < \nu/{\rm mHz} < 1/\delta t\sim90$, 
$\cal F$ shows many peaks of
comparable amplitude which on their own 
have little statistical significance.
However the coincidence in Figs.~\ref{fig_KPD2109} and \ref{fig_PB8783}
of the highest peaks in $\cal F$
with the highest peaks in $\cal P$  is 
more remarkable and is regarded as a positive signature of 
oscillatory behaviour in the spectroscopic data. 
The resolution $2/n\delta t\sim 0.1$mHz in $\cal F$ is insufficient to 
resolve the fine structure observed in much longer photometric time series.

%
%	Table                    Amplitudes of nrp modes in KPD\,2109+4401
%
\begin{table}
\caption[KPD2109 peak amplitudes]
{Velocity amplitudes of detected pulsation modes in KPS\,2109+4401. 
$\nu_p$ is the photometric frequency, 
$a_p$ the photometric amplitude (SDBV\,IX), 
$\nu_v$ the frequency of the measured velocity peak, 
$a_v$ the measured velocity amplitude. 
 }
\label{tab_KPD2109}
\small
\begin{center}
\begin{tabular}{clrr}
$\nu_p$ & $a_p$          & $\nu_v$ & $a_v$    \\
  mHz   & mmag           &  mHz    & \kmsec   \\[1mm]
 5.0455 &  4.4 $\rceil$  &         &          \\
 5.0842 &  0.9 $\,\vert$ &  5.09   &  2.49    \\
 5.0939 &  6.3 $\rfloor$ &         &          \\
 5.2124 &  2.6           &         &          \\
 5.4127 &  3.6 $\rceil$  &         &          \\
 5.4136 &  2.7 $\,\vert$   &  5.48   &  2.68    \\ 
 5.4818 &  4.9 $\rfloor$ &         &          \\
\end{tabular}
\end{center}
\end{table}

%
%    Table                           Multifrequency experiment
%
 \begin{table}
 \caption[Simulated multifrequency solution]
 {Simulated multifrequency observations: $\nu, a$ are the
 frequencies and amplitudes of the test signal, $n$ is the
 number of test runs, $\sigma$ is the standard deviation of
 normally-distributed noise. $a'$ are the
 mean and variance of the highest detected signal within
 $\pm0.1$ mHz of the principal test frequencies.}
 \label{tab_sim}
 \small
 \begin{center}
 \begin{tabular}{rlrrcc}
%% %\hline
 $\nu$     & $a$	& $n$ & $\sigma$ & $\langle a' \rangle $ & $a/a'$  \\
  mHz    &  \kmsec  &	& \kmsec & \kmsec  \\[2mm]

\multicolumn{6}{l}{KPD\,2109+4401 6 frequency simulation}\\[1mm]
  5.050 & 2.20 $\rceil$  & 100 & 15 &               &               \\
  5.080 & 0.45 $\,\vert$ &     &    &               &               \\
  5.090 & 3.15 $\rfloor$ &     &    & $3.23\pm0.76$ & $0.98\pm0.23$ \\
  5.410 & 1.80 $\rceil$  &     &    &               &               \\
  5.480 & 2.45 $\rfloor$ &     &    & $2.71\pm0.27$ & $1.06\pm0.08$ \\
  5.210 & 1.30           &     &    & $1.84\pm0.18$ & $0.71\pm0.10$ \\[2mm]

\multicolumn{6}{l}{KPD\,2109+4401 low-noise simulation}\\[1mm]
  5.050 & 2.20 $\rceil$  &  30 &  0 &               &               \\
  5.080 & 0.45 $\,\vert$ &     &    &               &               \\
  5.090 & 3.15 $\rfloor$ &     &    & $3.47\pm0.91$ & $0.91\pm0.26$ \\
  5.410 & 1.80 $\rceil$  &     &    &               &               \\
  5.480 & 2.45 $\rfloor$ &     &    & $2.53\pm0.10$ & $0.97\pm0.04$ \\
  5.210 & 1.30           &     &    & $1.48\pm0.05$ & $0.88\pm0.04$ \\[2mm]

\multicolumn{6}{l}{KPD\,2109+4401 2 frequency simulation}\\[1mm]
  5.090 & 3.15 & 100 & 15 & $3.09 \pm 0.31$ & $1.02\pm0.10$ \\
  5.480 & 2.45 &     &	  & $2.38 \pm 0.39$ & $1.03\pm0.16$ \\[2mm]
\multicolumn{6}{l}{PB\,8783 5 frequency simulation}\\[1mm]
  7.340 & 1.30 & 100 & 15 & $1.54 \pm 0.17$ & $0.84\pm0.11$ \\
  7.456 & 1.80 &     &    & $1.96 \pm 0.30$ & $0.92\pm0.15$ \\
  7.862 & 1.40 &     &    & $1.71 \pm 0.22$ & $0.82\pm0.13$ \\
  8.140 & 2.40 &     &    & $2.35 \pm 0.27$ & $1.02\pm0.12$ \\
 10.619 & 2.60 &     &    & $2.48 \pm 0.31$ & $1.05\pm0.13$ \\
\end{tabular}
\end{center}
\end{table}

\subsection{KPD\,2109+4401}

Because its spectrum is not contaminated by that of
a late-type companion, the results for KPD\,2109+4401 are the 
less complicated to interpret. 
A background drift of approximately 9\kmsec\ over 5 hours
is approximately linear and defines $b$. We adopted 
$\delta v=150\kmsec$ to establish $v_p(t)$ and to construct
Fig.~\ref{fig_KPD2109}.
Principal peaks in the amplitude spectrum are evident 
at $\nu_1 = 5.09$ mHz and $\nu_2 = 5.48$ mHz.

The same frequency structure in $\cal F$ was obtained for
$\delta v=150, 300$ and 450\kmsec\ when either a gaussian, parabolic
fit or centroid is used to measure $v(t)$.   In the interval 
4.5--6.4 mHz, minor peaks occured at $\nu_{3-7}=4.77$, 5.24, 5.33, 5.90 and 
6.03 mHz in most of these analyses.  
Beating may be responsible for some peaks; 
for example $\nu_3 = \nu_1-(\nu_2-\nu_1),
\nu_6 = \nu_2+(\nu_2-\nu_1)$. Thus only $\nu_1$ and $\nu_2$ could be
identified as genuine from the present data. 

The probability of finding a set of peaks at prespecified frequencies and
exceeding a given amplitude threshold by chance may be computed as follows.
For a given frequency range $\Delta v$, count the number of peaks $n$ 
which exceed a given threshold $v_{\rm thresh}$. Then count the number of
peaks $n_{\rm id}$ which lie within a resolution width $\pm \delta v$ of
the predicted frequencies. The probability of this number occurring by chance
is
\[
p = \biggl( n \frac{2\delta v}{\Delta v} \biggr) ^{n_{\rm id}}.
\]
For KPD\,2109+4401, we chose $\Delta v=6$\,mHz, $v_{\rm thresh}=1.5\kmsec$ and
$\delta v=0.05$\,mHz which gave $n=3$, $n_{\rm id}=2$ and hence $p=0.0025$.

In order to interpret the amplitudes of the identified peaks, 
we carried out the following numerical experiment (table~\ref{tab_sim}).  
We first formed the sum of a 
series of sine functions each having a frequency $\nu_{\rm i}$,
amplitude  $a_{\rm i}$ and a random phase shift $\phi_{\rm i}$. 
The $\nu_{\rm i}$ and $a_{\rm i}$ were based on the photometric results
for KPD\,2109+4401 (SDBV\,IX).
This sum was sampled at the same times $t_{\rm j}$ as the observations,
\[
s(t_{\rm j}) = \Sigma_{\rm i} a_{\rm i} \sin (2\pi \nu_{\rm i} t_{\rm j} + \phi_{\rm i} ).
\]
To this we added normally-distributed noise with a standard
deviation $\sigma=15\kmsec$ similar to the observational scatter.  
 $\cal F$ was derived and usually found
to contain peaks at  the principal input frequencies, but not always with 
the same amplitude ratios. Minor peaks were less frequently identified. 
The amplitudes
of the highest peaks within $\pm0.1$mHz of the test frequencies were
 measured.
This procedure was repeated $n$ times and the average 
amplitudes for each detected frequency $\langle a'_{\rm i}\rangle$ 
were computed. 
In the case of KPD\,2109+4401, the spacing of some input frequencies is less
than the resolution provided by the limited time series. Consequently, the
measured amplitudes represent some sum from several independent sine functions.
Interference, or beating, means that the apparent amplitude of such a 
signal will vary according to the relative phases of the components.  
This is apparent in table~\ref{tab_sim} (six-frequency simulation), 
where the relative standard deviation of the amplitude of the 
composite signal at 5.09\,mHz is much higher 
than that of the single-frequency signal at 5.21\,mHz.
The experiment was repeated with only two well-resolved input frequencies 
and yielded the ratio $ a'/a \sim 1.0\pm0.1 $. 
Reducing $\sigma$ reduces the error on $a'/a$ due to experimental noise,
but not that due to mode beating or to data sampling.
It was also repeated for 
$\nu$ and $a$ appropriate to the second target PB\,8783 with a similar result.
 
% Kjeldsen \& Bedding (1993; Eqns. A3, A4) performed similar 
% simulations to determine the effect of noise ($\sigma_{\rm amp}$) on the 
% measurement of amplitudes in power spectra. In contrast to our 
% simulations, they obtained
% \[
% a_1^2 = a_{\rm osc}^2 + (8.8\pm2.3) \sigma_{\rm amp}^2 
% \]
% \[
% a_5^2 = 0.94a_{\rm osc}^2 + (3.4\pm1.1) \sigma_{\rm amp}^2 
% \]
% for the measured amplitudes of a single mode and the average of five modes 
% respectively. However, their simulations involved $\sim20$ input frequencies
% with regular spacings and a continuous time series so that constructive 
% interference with noise peaks, for example, significantly affects the result.

Therefore, we conclude that the peak amplitudes of 2.49 and 2.68 \kmsec\ 
at frequencies 5.09 and 5.48 mHz in KPD\,2109+4401 reflect to within 
$\sim20\%$  (standard deviation) the combined real amplitudes of oscillations 
within $\sim0.05$mHz of these frequencies  (Table~\ref{tab_KPD2109}).
The principal causes of uncertainty in the amplitudes are beating between
unresolved modes and experimental noise. 

The 9 \kmsec\ drift in $b$ may be real. Instrumental drift is limited to
$\simle0.9 \kmsec$ (section 2). The heliocentric correction, which was not applied 
to the data, changes by less than 0.4\kmsec. In the simulations described above, 
the r.m.s. amplitude of $b$ was 1.7\kmsec. Although KPD\,2109+4401 is not known 
to be a binary, it cannot be ruled out. PG\,1336--018 was only recognised as a 
binary from its eclipses; the secondary is too faint to be detected spectroscopically 
(SDBV\,VIII).

\subsection{Line profile variations?}

In addition to radial-velocity information, $\chi$ contains information 
about line-profile variations, since the ccf represents an average line
profile within the region of spectrum analyzed. In the present case, 
the average line profile is dominated by two Balmer lines, H$\gamma$ and H$\delta$.
The average line width and asymmetry are indicated either by the 
moments $M_{2,3}$, respectively, or by the coefficients $g_{2,4}$ in the 
gaussian + linear fit 
$g(v)=g_0 \exp ( (v-g_1)^2/g_2 ) + g_3 + g_4 v$. 

Neither the Fourier
transform ${\cal F} \{ M_2(t)\} $ or ${\cal F}\{ g_2(t)\}$  showed 
peaks near $\nu_1$ or $\nu_2$. With suitably chosen $\delta v$,
both ${\cal F} \{ M_3(t)\} $ and  ${\cal F}\{ g_4(t)\}$ showed frequency 
structure similar to ${\cal F}\{ v(t)\}$. However, the amplitude of
peaks in ${\cal F} \{ M_1(t)\} $ only approached that of ${\cal F}\{ v_p(t) \}$
if $\delta v$ was sufficiently large, in which case the periodic content of
$M_3(t)$ was essentially lost. This implied that $M_3$ contained velocity
information not properly incorporated into $M_1$.  

${\cal F}\{ g_4(t)\}$ showed
peaks at 5.09 and 5.55 mHz with $\delta v \sim 400-600 \kmsec$, but not otherwise. 
In the case of intrinsically sharp lines, such asymmetry may reflect 
differential radial motion across the rotationally broadened line profile.
In the present study, any rotational broadening is convolved with 
Stark broadening in the Balmer lines and is difficult to interpret. 
Consequently, whilst we suspect that there may be evidence for line-profile
variations with similar frequencies to the radial motions, we can
draw no firm conclusions from the present data.

%
% Figure                                            PB8783 - ccf
%
\begin{figure}
\epsfxsize=85mm
\epsfbox{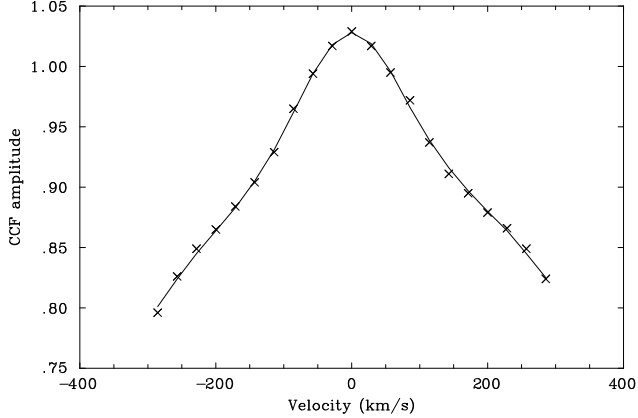}
\caption[PB8783: typical ccf]{Peak of a typical cross-correlation
function $\chi(v)$ for PB\,8783 (crosses). The gaussian + quadratic
fit used to construct the two-component velocity solution 
(Figs.~\ref{fig_rvB} and \ref{fig_rvF}) is also shown (solid line).}   
\label{fig_ccf}
\end{figure}

%
%	Table				    Amplitudes of nrp modes in PB\,8783
%
\begin{table}
\caption[PB8783 peak amplitudes]
{Velocity amplitudes of detected pulsation modes in PB\,8783. Symbols are
as in table~\ref{tab_KPD2109} except that $\nu_p, a_p$ are from SDBV\,V. 
The identification marked (:) is tentative.}
\label{tab_PB8783}
\small
\begin{center}
\begin{tabular}{clrr}
$\nu_p$ & $a_p$           & $\nu_v$ & $a_v$  \\
  mHz   &  mmi            &   mHz   & \kmsec \\[1mm]
 7.338  &  0.28           &  7.339  &  1.22  \\
 7.453  &  1.78 $\rceil$  &         &        \\
 7.454  &  1.09 $\rfloor$ &  7.456  &  1.54  \\
 7.870  &  1.62           &  7.857  &  1.23  \\
 8.090  &  1.20           &         &        \\
 8.1507 &  2.09 $\rceil$  &         &        \\
 8.1516 &  1.65 $\,\vert$ &  8.141  &  1.91  \\
 8.1526 &  1.18 $\rfloor$ &         &        \\
 8.589  &  0.27           &  8.561: &  1.09: \\
10.587  &  0.17 $\rceil$  &         &        \\
10.623  &  0.38 $\rfloor$ & 10.625  &  2.18  \\
\end{tabular}
\end{center}
\end{table}

\subsection{PB\,8783}

The binary PB\,8783 is altogether more difficult to analyze because 
of the  superposition of sdB and F star spectra. The results shown in
Fig.~\ref{fig_PB8783} are based on fitting parabola to $\chi(v)$ 
with $\delta v = 420\kmsec$. 
In deriving the amplitude spectrum in Fig.~\ref{fig_PB8783}, the background 
function $b(t)$ was found to be best given by a sinusoid with an amplitude 
of 10\kmsec, representing a systemic drift substantially larger than that 
in KPD\,2109+4401.

On inspection, the typical ccf for PB\,8783 consists 
of a strong broad component (FWHM $\sim 1400 \kmsec$)
superimposed by a small narrow component (FWHM  $\sim 200 \kmsec$, 
Fig.~\ref{fig_ccf}).
The narrow feature disappears when the template is
replaced by a theoretical sdB star spectrum containing H and He lines only,
and is deduced to be due principally to metal-lines
in the F-type secondary. This is confirmed when $\chi$ is calculated
using only wavelengths between H$\delta$ and H$\gamma$; the broad component
disappears leaving a small narrow peak. 

The detection of two components in 
$\chi$ offers the possibility to 
resolve the radial motions of the binary components. The
mutual motion of the F-star is clearly apparent when $\chi$ excludes
the Balmer lines, and appears as a change of $\sim10\kmsec$ during the
five-hour observing run. However, the sign of the change remains the
same when the theoretical sdB spectrum is used as a template, 
indicating that the Balmer lines in the F-star contribute significantly 
to $\chi$. 
 
%
% Figure                                            PB8783 - blue-components
%
\begin{figure}
\epsfxsize=85mm
\epsfbox{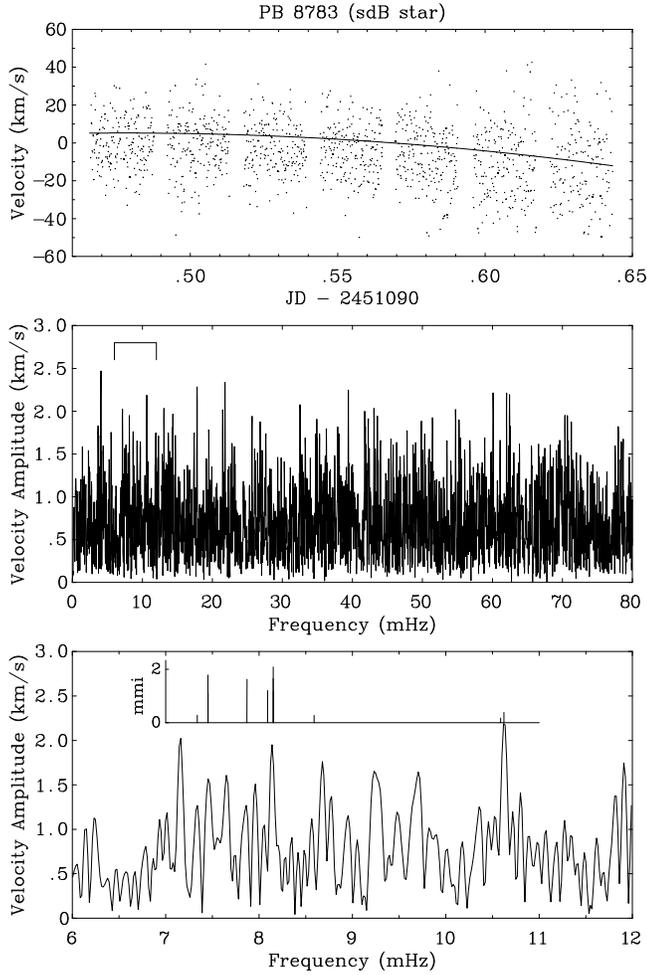}
\caption[PB8783: hot component]{Radial velocities and amplitude 
spectrum of the sdB star in PB\,8783. The least squares solution to the
slowly varying velocity component is shown in the top panel (solid line).
The amplitudes of photometric frequencies are as in Fig.~\ref{fig_PB8783}.
}   
\label{fig_rvB}
\end{figure}

%
% Figure                                            PB8783 - red-components
%
\begin{figure}
\epsfxsize=85mm
\epsfbox{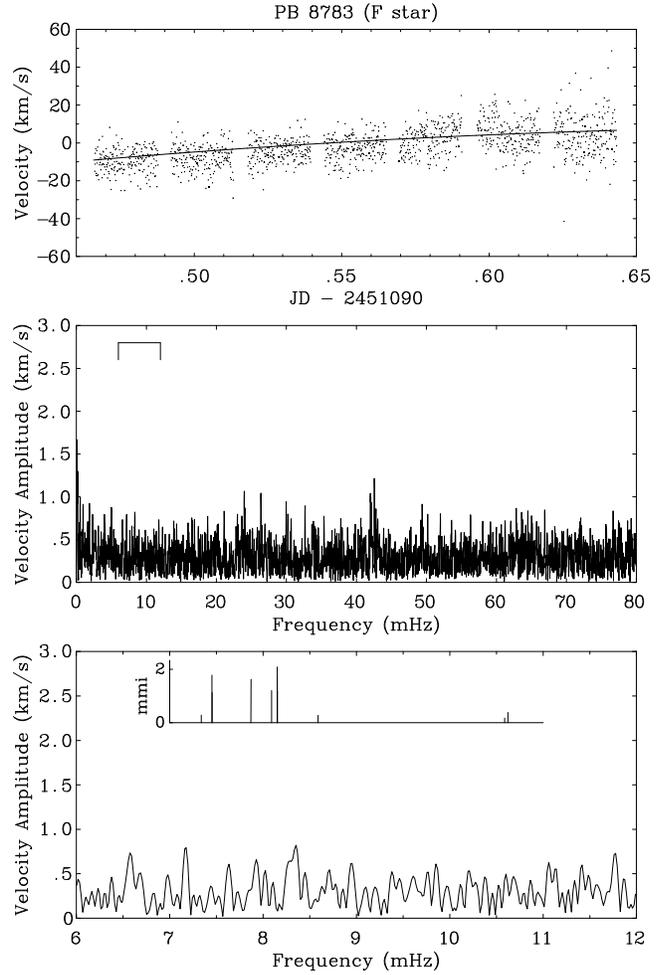}
\caption[PB8783: cool component]{As Fig.~\ref{fig_rvB} for the F star in PB\,8783.
}   
\label{fig_rvF}
\end{figure}

The problem was resolved by fitting 
$\chi(-\delta v:+\delta v), \delta v = 280\kmsec$ with the function
$g(v)=g_0 \exp ( (v-g_1)^2/g_2 ) + g_3 + g_4 v + g_5 v^2$, where the
quadratic terms fit the broad component and the 
gaussian terms represent the narrow component (Fig.~\ref{fig_ccf}). 
The functions
$g_1(v)$ and $-g_4(v)/2g_5(v)$ were examined to determine the mutual motion of
each star and to look for high-frequency content.
The results are
shown in Figs.~\ref{fig_rvB} and \ref{fig_rvF} where it will be seen that the 
slowly varying components are of opposite sign. It is suggested that, in the absence
of viable alternatives for these systematic drifts, these slowly varying components
are most probably due to the mutual motion of the two stars within the binary. 

If this is the case, a linear regression on the first 4 hours of both sets of 
velocity data gives the relative acceleration of the two components. The result,
$\Delta v_{\rm F}/\Delta v_{\rm sdB}=-0.81\pm0.10$, 
provides an estimate for the mass ratio of the two stars
$M_{\rm sdB}/M_{\rm F}\sim0.8$. This is a little high for an F-star
on the main-sequence ($1.1-1.8\Msolar$) and a canonical 
sdB star ($\sim0.5\Msolar$), and suggests that the relative motions have 
still not
been fully resolved.  Further information concerning 
the velocities of the  two components relative to one another is convolved 
into the composite Balmer profiles in the ccf template and will 
require observations of the complete binary orbit to disentangle.
Whilst the binary period is clearly longer than the data sequence, it
is constrained by the relation
\begin{equation}
K_{\rm F} = 210 \frac{1}{1+q} \left( \frac{M}{P} \right)^{1/3} \sin i \kmsec.
\end{equation}
$M$ is the total mass of the system ($\Msolar$), $q=M_{\rm F}/M_{\rm sdB}$
is the mass ratio, $P$ is the orbital period (d) and $K_{\rm F}$ is the
velocity semi-amplitude of the F star (\kmsec). For example, with  
$M_{\rm sdB}=0.55\Msolar$, $M_{\rm F}=1.2\Msolar$, $i=30^{\circ}$ and $P=2$d, 
$K_{\rm F} = 32\kmsec$. The maximum change in velocity over the
timescale of our observations (0.15d) would be 
$\Delta v_{\rm F}\sim15 \kmsec$, compared with 13\kmsec\ observed. 
Inverting the procedure, we can determine the maximum $P$ as a function of $i$ 
required to obtain $\Delta v_{\rm F}=13 \kmsec$ in 0.15d. For 
$M_{\rm F}=1.2-1.8\Msolar$, the maximum $P$ increases from $1.0-0.8$d at
$i=10^{\circ}$ to $P=3.7-3.2$d at $i=80^{\circ}$. Eclipses are not expected at 
inclinations $\simle80^{\circ}$. 

It is satisfying to detect the photometric frequencies already reported
for PB8783 (SDB\,V) in  the deconvolved radial velocities of the sdB star
(Fig.~\ref{fig_rvB}) and to find no such signal in the radial 
velocities of the F star (Fig.~\ref{fig_rvF}). This provides confirmation that
the photometric and radial velocity variations are due to pulsations
in the sdB star and not in the cool companion.  A similar
result was obtained for PG1336-018 (SDBV\,VIII), except in that case the 
persistence of oscillations during an eclipse of the unseen companion
provided the necessary proof. 

From this point, the interpretation of 
the sdB star velocities is the same as that for KPD2209+4101. 
Peaks in the velocity amplitude spectrum were identified  at 
five frequencies also present in the photometry. 
The resulting pulsation velocity amplitudes were thus found to be in the 
range $1.1-2.2\kmsec$ and are shown in Table~\ref{tab_PB8783}.
Regarding the probability that these frequencies were identified by
chance and following the prescription already given for KPD\,2209+4101,
we found with $v_{\rm thresh}=1.5\kmsec$ that 
$n=9, n_{\rm id}=3$ and $p=0.0034$ (Fig.~\ref{fig_rvB}).
Relaxing $v_{\rm thresh}=1.0\kmsec$ yielded $n=27, n_{\rm id}=6$ and 
$p=0.0083$.

\subsection{Relating luminosity and velocity amplitudes}

Kjeldsen \& Bedding (1995) investigated the relation between velocity
and luminosity amplitudes for several classes of main-sequence,
giant and supergiant pulsators and derived the calibrated relation
\begin{equation}
 (\delta L/L)_{\lambda} = \frac{v_{\rm osc}/ {\rm m\,s^{-1}}}
    { (\lambda / 550 {\rm nm}) (\Teff/5777 K)^2 } 20.1 {\rm ppm}, \label{eq_dl} 
\end{equation}
where ppm denotes parts-per-million. 

We tested whether these predictions could be extended to the sdBVs since we 
have both light ($\delta L/L$) and velocity ($v_{\rm osc}$) 
amplitudes for the same oscillation frequencies. 
We found that the observed velocity amplitudes are $>2$ times
those expected from the photometry and Eqn.~(\ref{eq_dl}). 
However Eqn.~(\ref{eq_dl}) has been derived from cool stars pulsating in
radial or non-radial p-modes and it may not be applicable to 
pulsations in the hot sdBVs.  We note that   
the amplitudes of individual modes in sdBVs have been reported to vary from
one season and possibly one night to another (SDBV/,VIII). Beating
between unresolved modes will certainly affect the velocity amplitude
measurements. Simultaneous
high-speed photometry and spectrocopy will be required to properly relate
the velocity and luminosity amplitudes and to make any futher
validation of  Eqn.~\ref{eq_dl}.

\subsection{For future investigation}

We have left two issues for future investigation. \\
1) Compared with its apparent surface gravity, the observed pulsation 
frequencies of PB\,8783 are too high (SDBV\,XII). The expectation is
that its surface gravity ($\log g=5.55$, O'Donoghue et al. 1997, SDBV\,IV) should be much higher. 
It is recognised that this could be due to inadequate subtraction of 
the F-star spectrum, a conjecture which should be checked using 
higher quality data, including those presented here. \\
2) If the mode and amplitude of a spherical harmonic responsible for 
each pulsation frequency is specified, then it is relatively straightforward 
to integrate the projected surface velocities to provide a theoretical
radial velocity curve. It is slightly more complicated to construct the
theoretical light curve. From the combination it should be
possible to refine the mode and amplitude measurements already established. 

\section{Conclusions}

High-speed high-resolution spectroscopy of two non-radially pulsating 
subdwarf B stars KPD\,2109+4401 and PB\,8783 has been acquired with 
the William Herschel telescope.
These data show radial velocity variations at both high- and low-frequencies.
High-frequency peaks in the velocity amplitude spectrum correspond to
frequencies identified in the light curves of both stars, and have allowed
the velocity amplitudes associated with these stellar oscillations to be 
estimated. Typically $\sim2\kmsec$, these translate into radial
variations of some $100-250$km within 60--90s. The prospect for 
measuring the velocity amplitudes of non-radial pulsations in sdBVs
with much higher amplitudes is promising. Line profile variations 
were not detected in the data although there are good reasons to 
suppose they should be present. The possibility that frequencies not present 
in the photometry may contribute to the velocity amplitude spectrum
should also be pursued.

In the case of the binary sdB PB\,8783, low-frequency velocity variations 
corresponding to the mutual motion of both components have been resolved.
These provide an upper limit to the orbital period of between 0.8 and 3d,
depending principally on the orbital inclination. Establishing the orbital
period for this star should therefore be a priority for future
observations. The deconvolution of component star velocities also demonstrated 
that high-frequency velocity variations occur only in the sdB star 
and not in the F star.

%% \section{Conjectures}

\section*{Acknowledgments}

This research has been supported by the Department of Education in 
Northern Ireland through a grant to the Armagh Observatory.

%
% _________________________________________________________________ references
%

\label{lastpage}

\end{document}